\documentclass[12pt]{article}
\usepackage{epsf}
\usepackage{pazh}
\tightenlines

\sloppypar

\begin{document}
\baselineskip 21pt

ÓÄÊ 524.354.4

\title{\bf CLASSICAL CEPHEIDS AND THE SPIRAL STRUCTURE OF THE MILKY WAY}

\author{\bf \hspace{-1.3cm}\copyright\, 2015 \ \ 
A.K. Dambis\affilmark{1*},  L.N.Berdnikov\affilmark{1,2}, Yu.N.Efremov\affilmark{1},
A.Yu.Kniazev\affilmark{1,3,4}, A.S.Rastorguev\affilmark{1}, E.V.Glushkova\affilmark{1},
V.V.Kravtsov\affilmark{1,5}, D.G.Turner\affilmark{6}, D.J.Majaess\affilmark{6},
R.Sefako\affilmark{3}}

\affil{
{\it Sternberg Astronomical Institute, Lomonosov Moscow State University, 
Universitetskii pr. 13, Moscow, 119992 Russia}$^1$\\ 
{\it Astronomy and Astrophysics Research division, Entoto Observatory and Research Center, P.O.Box 8412, Addis Ababa,
Ethiopia}$^2$
{\it South African Astronomical Observatory, P.O. Box 9, Observatory, Cape Town, 7935, South Africa}$^3$
{\it Southern African Large Telescope, P.O. Box 9, Observatory, Cape Town, 7935, South Africa}$^4$
{\it Departamento de F\'isica, Faculdad de Ciencias Naturales, Universidad de Atacama, Copayapu 485, Copiap\'o, Chile}$^5$
{\it Department of Astronomy and Physics, Saint Mary's University, Halifax, NS B3H 3C3, Canada}$^6$
}

\vspace{2mm}
\received{\today}

\sloppypar \vspace{2mm} \noindent 
We use the currently most complete  collection of reliable Cepheid positions (565 stars) out to 
$\sim$~5 kpc based mostly on our photometric data to outline the spiral pattern of our Galaxy.  
We find the pitch-angle to be equal to 9--10$^{\rm o}$ with the most accurate estimate 
($i$~=9.5~$\pm$0.1$^o$) obtained assuming that the spiral pattern has a four-armed structure,
and the solar phase angle in the spiral pattern to be $\chi_{\odot}$~=~121$^o$~$\pm$~3$^o$. 
The pattern speed 
is found to be $\Omega_{sp}$~=~25.2~$\pm$~0.5~km/s/kpc based on a comparison of
the positions of the spiral arms delineated by Cepheids and maser sources and the
age difference between these objects. 



\vfill
\noindent\rule{8cm}{1pt}\\
{$^*$ E-mail: $<$mirage@sai.msu.ru$>$}

\clearpage

\section*{INTRODUCTION}
\noindent Positions of spiral arms in our Galaxy remain a topic
of active debate. Thus H~I data consistently indicate the existence
of two -- the Carina and Cygnus (or Outer) -- arms (Nakanishi \& Sofue 2003; Levine,
Blitz \& Heiles 2006). Englmaier \& Gerhard (1999),  Englmaier, Pohl, \& Bissantz (2011),
and Hou, Han, \& Shi (2009) pointed out that the four-armed spiral pattern fits well the distribution of 
neutral (H~I), molecular ($H_2$), and ionized (H~II) hydrogen;  at the same time,  Pohl, Englmaier, 
\& Bissantz (2008) proposed a two-armed model based on the results of the simulation
of gas streaming motions. Russeil (2008), in turn, constructed a four-armed spiral pattern based on the 
distribution of  star-forming regions; Efremov (2011) also pointed out that the data for molecular and 
neutral hydrogen regions are also quite consistent with the four-armed spiral structure whose fragments
include the most pronounced Carina and Cygnus arms, where H~I and $H_2$ superclouds are found to 
be spaced at regular intervals. A more detailed picture can be recovered from the distribution
of stars  known to concentrate towards spiral arms. 

Observations show that in other spiral galaxies gas and young stellar populations concentrate toward
practically the same arms and hence young stars and clusters should be good tracers of the
spiral pattern. However, we reside inside the Milky Way and therefore cannot look at it
from outside like an external galaxy. Hence our tracers have also to be good distance indicators 
observable at large enough distances to qualify for the task. 
Maser sources have become very promising
tracers of the structure and kinematics of the Galactic disk in recent years
because of their accurate radial velocities combined with precise VLBI trigonometric parallaxes and proper
motions determined within the framework of several dedicated projects (Kim et al. 2008;
Reid et al. 2009a, Rygl et al. 2010; Bobylev \& Bajkova, 2013). However, the number of masers with 
precisely measured parameters remains rather small (about 130 objects) to allow detailed tracing
of the Galactic spiral structure. Classical Cepheids, which are young stars whose relative distances
can be estimated very accurately via the period-luminosity relation,  appear to be among the objects 
best suited for the task. 

Thus recent studies of M31 resulted in the discovery
in that galaxy of more than 2000 Cepheids (Kodric et al.
2013), all of which, except for about 150 Type II Cepheids,
concentrate to the spiral arms, and we can expect Cepheids in our own Galaxy to
behave similarly. Extensive precise multicolor photometry collected for
Milky-Way Cepheids over many years (Berdnikov 2008; Berdnikov et
al. 2011, 2014), combined with a reliable period-luminosity
relation (Berdnikov, Dambis \& Voziakova 2000), make these variable stars efficient tracers 
of the positions of spiral arms out to a heliocentric distance of about 5 kpc,
allowing the parameters of the entire spiral structure to be
determined under the assumption of its regularity. 

\section*{OBSERVATIONAL DATA FOR THE MILKY WAY CEPHEIDS}
\noindent
Berdnikov el al. (2000) and Berdnikov (2006) published
the photometric parameters for practically all Galactic Cepheids from the General
Catalogue of Variable Stars (Samus et al. 2007) based
on extensive photometry obtained with
0.5-m to 1-m reflectors on Mt. Maidanak (Uzbekistan Republic),
Cerro Tololo and Las Campanas (Chile), and the
South African astronomical observatories. CCD monitoring
of the southern sky performed as part of the ASAS project
(Pojmanski 2002) resulted in the discovery of tens of thousands
of new variable stars, more than a thousand of which
were classified as Cepheids.

However, because of short focal lengths
of the cameras employed,  for the new Cepheids to
be usable as distance indicators they had to be reobserved
in several colors with ''normal'' telescopes. We therefore obtained
CCD and photoelectric observations of 164 classical
Cepheids from the ASAS-3 catalog during nine observing
runs between April 2005 and January 2013 (Berdnikov et al.
2009a, 2009b, 2011, 2013, 2015) with the 76-cm telescope of the
South African Astronomical Observatory (SAAO, Republic
of South Africa) and the 40-cm telescope of the Observatorio
Cerro Armazones (OCA) of the Universidad Catolica
del Norte (Chile). The observations were obtained with an
SBIG ST-10XME CCD camera and a single-channel pulsecounting
photoelectric photometer equipped with BVIc filters of the 
Kron-Cousins system (Cousins 1976).

In our study of the space distribution of classical Cepheids in the Milky Way we use 
the updated version of the catalog of Cepheid parameters by 
Berdnikov el al. (2000), which includes the data for 674 stars. 
The procedure of deriving distances is based on the K-band period-luminosity
relation of Berdnikov, Vozyakova \& Dambis (1996b),
interstellar-extinction law derived in Berdnikov, Vozyakova
\& Dambis (1996a), and the interstellar-extinction values estimated using the 
$B-V$ period-color relation of Dean, Warren, and Cousins (1978). Since the
underlying distance-scale calibration yields a solar Galactocentric distance
of $R_0$=7.1~kpc, we adopt it throughout this paper for consistency. Note that
our results expressed in terms of $\theta$ and $R_G/R_0$ 
remain practically unchanged if we compute distances based on a 
more recent PL relation of Fouque et al.~(2007) and at the same time upscale $R_0$ to 
$\sim$~8.0~kpc.

\section*{SPIRAL ARMS OUTLINED BY CEPHEIDS}
\begin{figure}[ht]
\epsfxsize=13cm
\hspace{-2cm}\epsffile{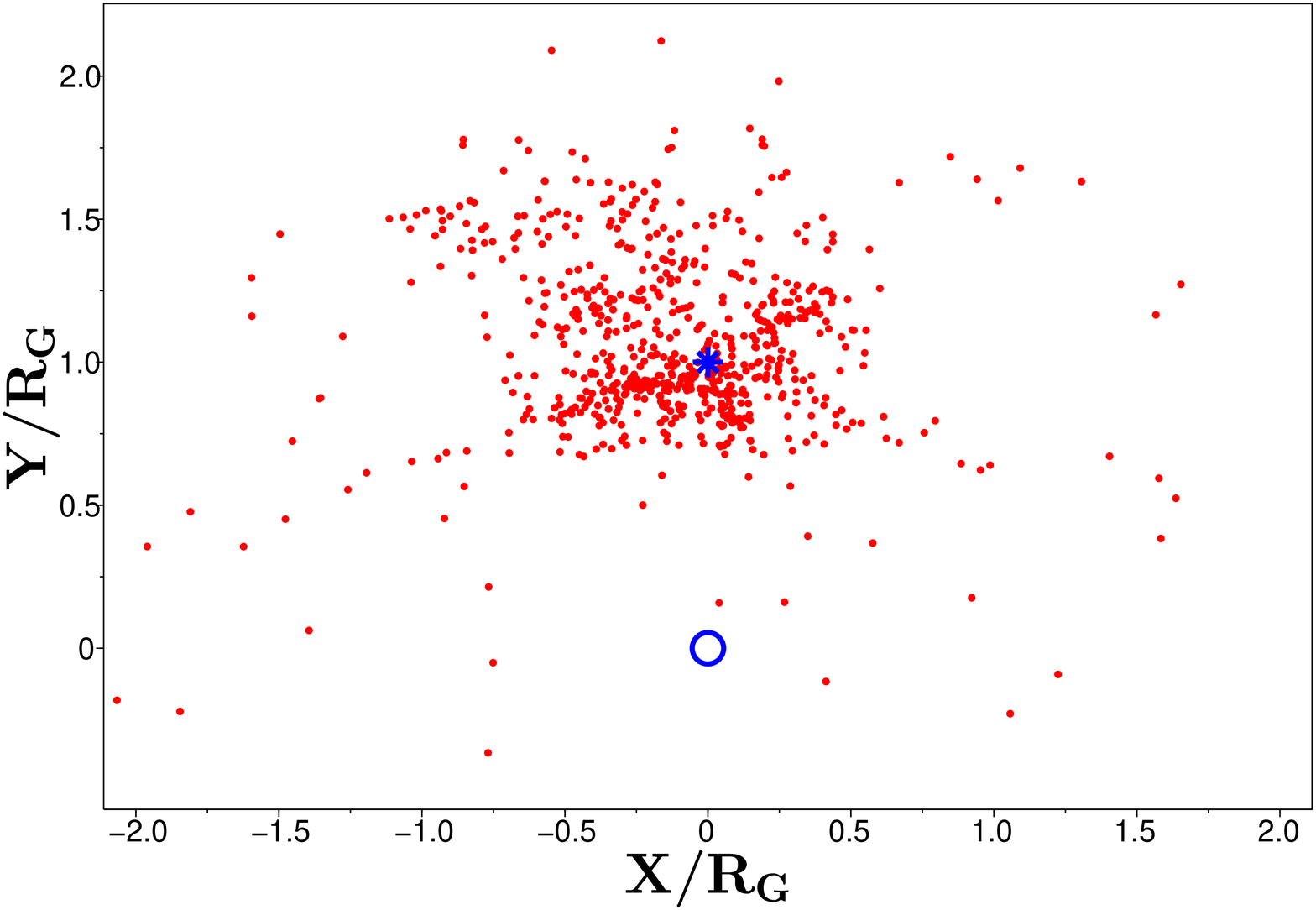}

\caption{\rm }
\label{xyall}
\end{figure}

\begin{figure}[ht]
\epsfxsize=13cm
\hspace{-2cm}\epsffile{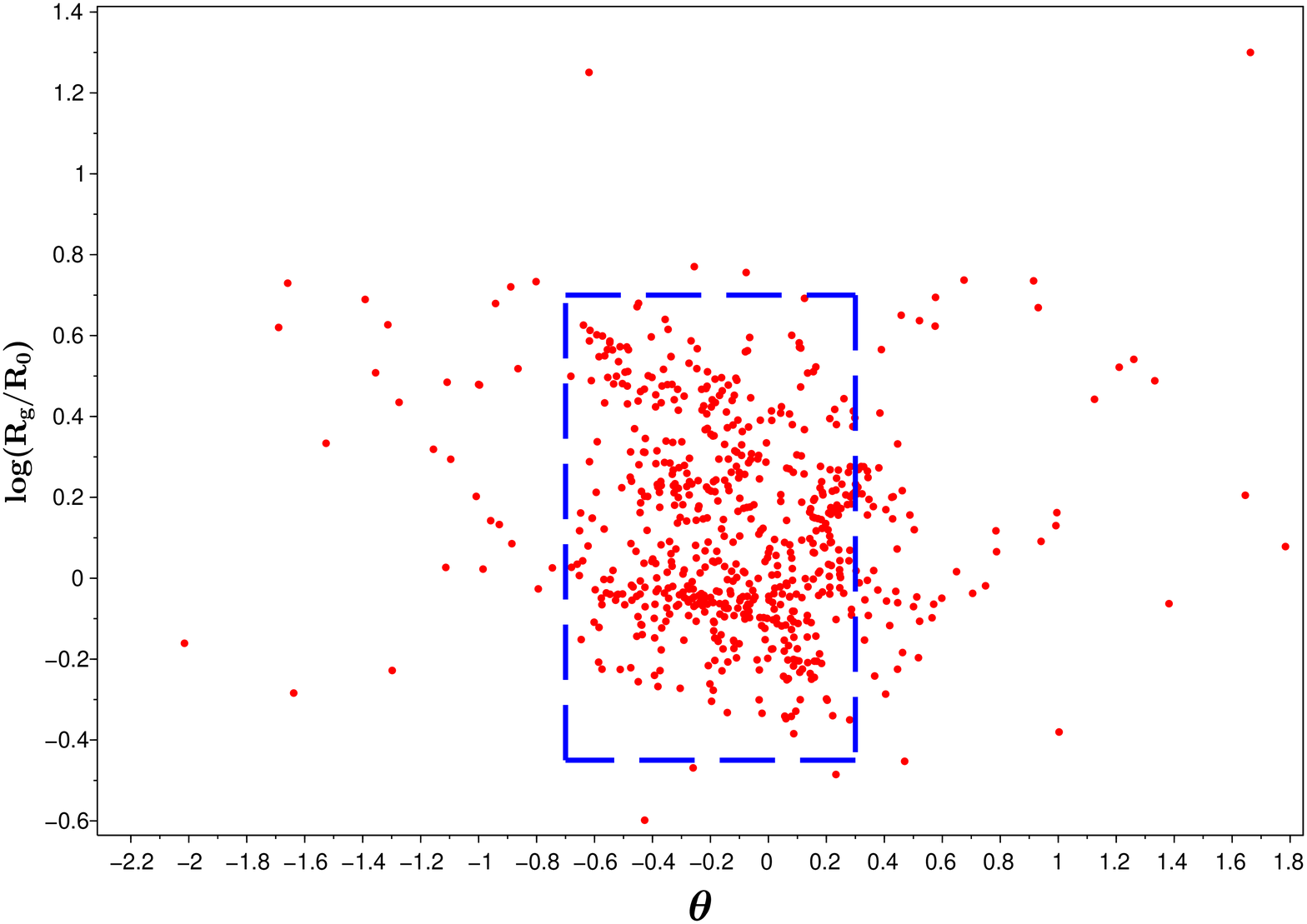}

\caption{\rm }
\label{thetalgrall}
\end{figure}

\noindent
Figure~\ref{xyall} shows the distribution of our Cepheid sample projected onto the 
Galactic plane. Figure~\ref{thetalgrall} shows the same Cepheids in the
log ($R_g/R_0$), versus Galactocentric azimuth, $\theta$, coordinates.
The latter diagram is more convenient for identifying eventual spiral patterns because logarithmic spirals
\begin{equation}
R_G/R_0=a_0 e^{(\theta-\theta_0)\tan i},
\label{spiral}
\end{equation}
become straight lines
\begin{equation}
log (R_G/R_0)= \alpha_0 +(\theta-\theta_0)\tan i
\label{lspiral}
\end{equation}
where $\alpha_0$ = log $a_0$, $i$ is the pitch angle, and $\theta_0$ is some arbitrary
initial angle, which we set equal to zero, $\theta_0$~=~0  (see, e.g., Bobylev \& Bajkova (2013)).
As is evident from this figure, the distribution is highly incomplete outside the region 
-0.7~$\leq$~$\theta$~$\leq$~+0.3, -0.45~$\leq$~log($R_g/R_0)$~$\leq$~+0.7 shown in Figs.~\ref{xyall} and
\ref{thetalgrall} by the dashed annular sector and dashed rectangle, respectively. Hereafter we 
use only the 565 stars located within this sector/rectangle in our search for spiral fragments.

\subsection*{Search for the major arm fragment}
\begin{figure}[ht]
\epsfxsize=13cm
\hspace{-2cm}\epsffile{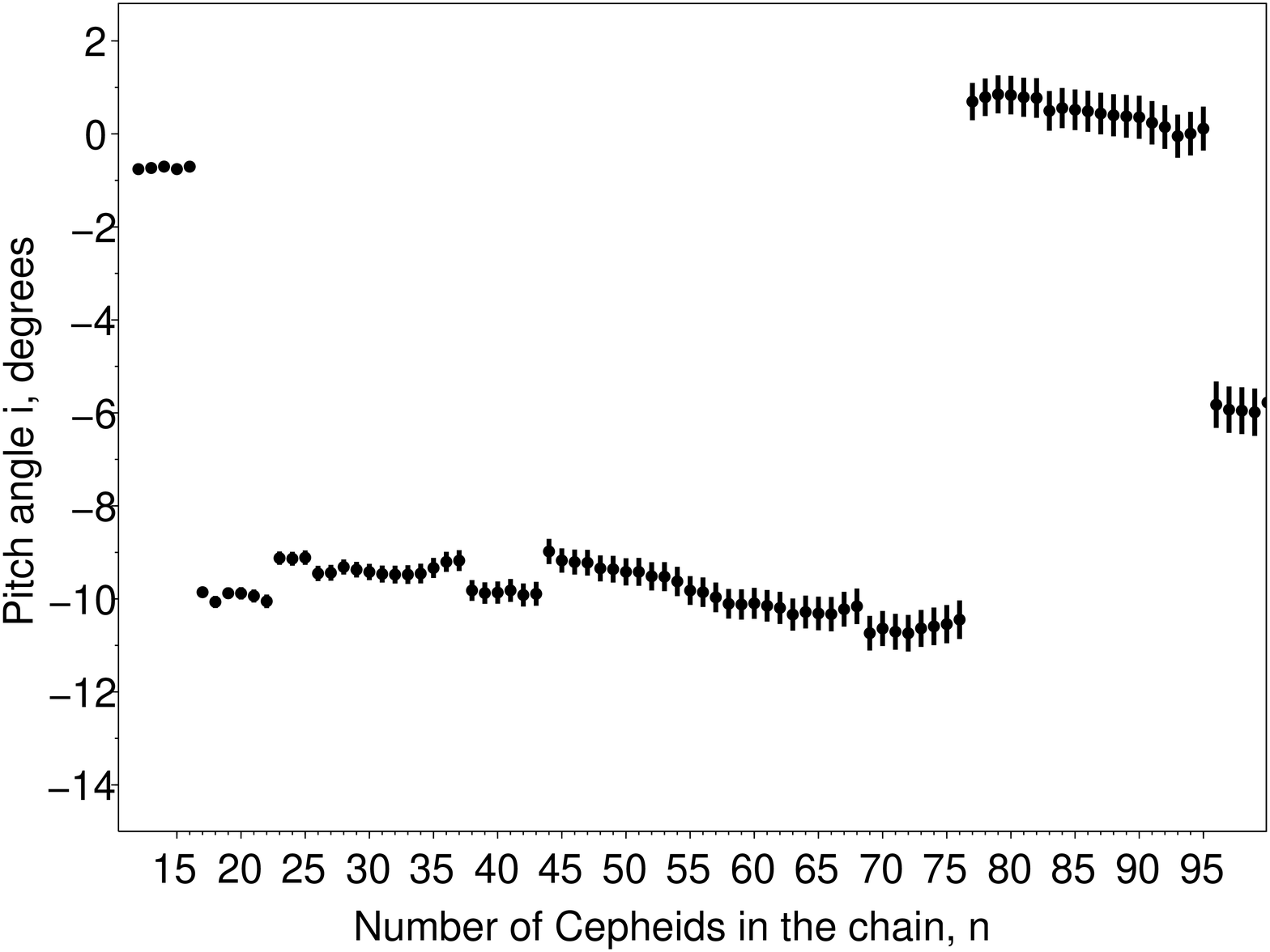}

\caption{\rm }
\label{spirpar1}
\end{figure}

\begin{figure}[ht]
\epsfxsize=13cm
\hspace{-2cm}\epsffile{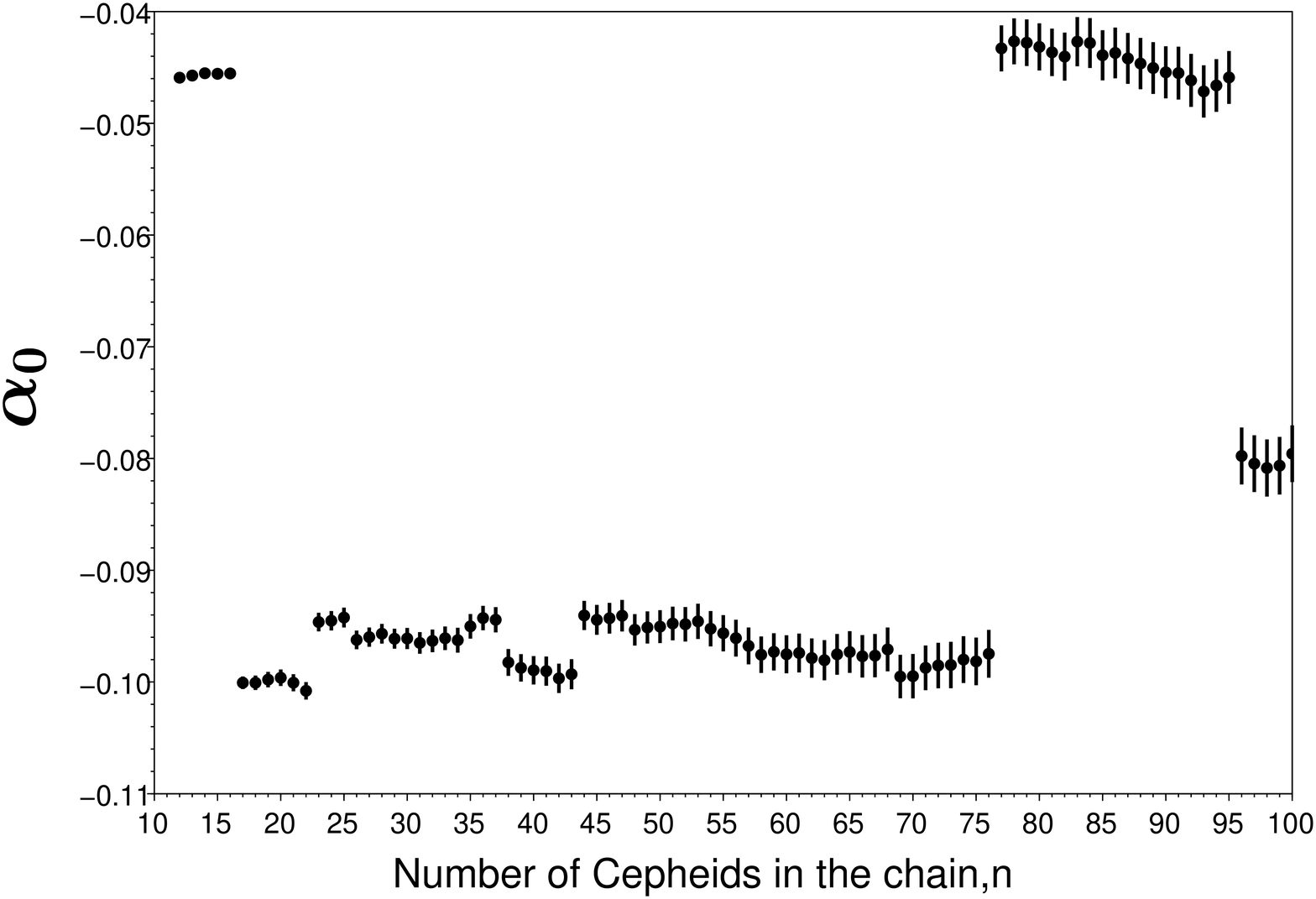}

\caption{\rm }
\label{spirpar2}
\end{figure}

\noindent
Given that spiral-arm equation is linear in the coordinates $\theta$ -- log~($R_G/R_0$), our task is 
to identify linear chains in Fig.~\ref{thetalgrall} that contain
as much  points as possible and at the same time have the smallest possible cross-chain scatter. 
To this end we use a slightly modified version of the ''simple approach'' described by Guerra 
\& Pascucci (2001).
The idea of this simple approach is ''to search for the optimal line among those defined by
pairs of points in the dataset''. In the version outlined in the above paper the number of other points 
lying within a certain ''distance'' $\epsilon$ from the two-point line defined is determined, and the
best line is considered to be the one with the maximum number of such points. 
Our approach differs in that instead of
fixing $\epsilon$ and maximizing $n$ we  fix the the number of stars $n$ and minimize
the cross-line scatter. More specifically, for
each two-point line we find $n$ points that are closest to it (including the two line-defining points), 
and determine the mean squared deviation of selected $n$ points from this line along the log~($R_G/R_0$) 
direction. 
Our best $n$-point line  is the one with the smallest mean squared deviation of $n$ points closest to it. 
To identify the dominating spiral in the distribution of Cepheids, 
we search for such best lines corresponding to various $n$ in the $n$~=~10--100 interval and 
see how the parameters of these lines ($\alpha_0$ and tan~$i$) vary with increasing $n$. 
In Figs.~\ref{spirpar1} and \ref{spirpar2} we plot the
resulting parameter estimates $\alpha_0(n)$ and $i(n)$ as a function of $n$ in the~10--100
interval for our sample of 565 Cepheids. 
As is evident from the figures, both parameters are rather stable in the range  
$n$~=17--76  ($\alpha_0(n)$ remains within the -0.094 to -0.101 interval,
and $i(n)$ within the 8.98$^o$--10.74$^o$ interval) changing abruptly with further increase of $n$. We 
therefore
assume that the chain of $n_{max}$=76 stars with the minimal scatter outlines the major spiral arm in the
distribution of Cepheids considered.  The parameters of the resulting spiral (determined
via linear least-squares solution based on $n_{max}$=76 stars) are:

\begin{equation}
\alpha_0 = -0.097 \pm 0.002~\hspace{0.3cm}~{\rm and}~{\rm tan}~i = -0.184 \pm 0.007~\hspace{0.3cm}~(i = -10.45^o \pm 0.41^o)
\label{carina_arm}
\end{equation}

with the cross-spiral scatter of $\sigma~{\rm log} (R_G/R_0) = 0.013$.

\begin{figure}[ht]
\epsfxsize=13cm
\hspace{-2cm}\epsffile{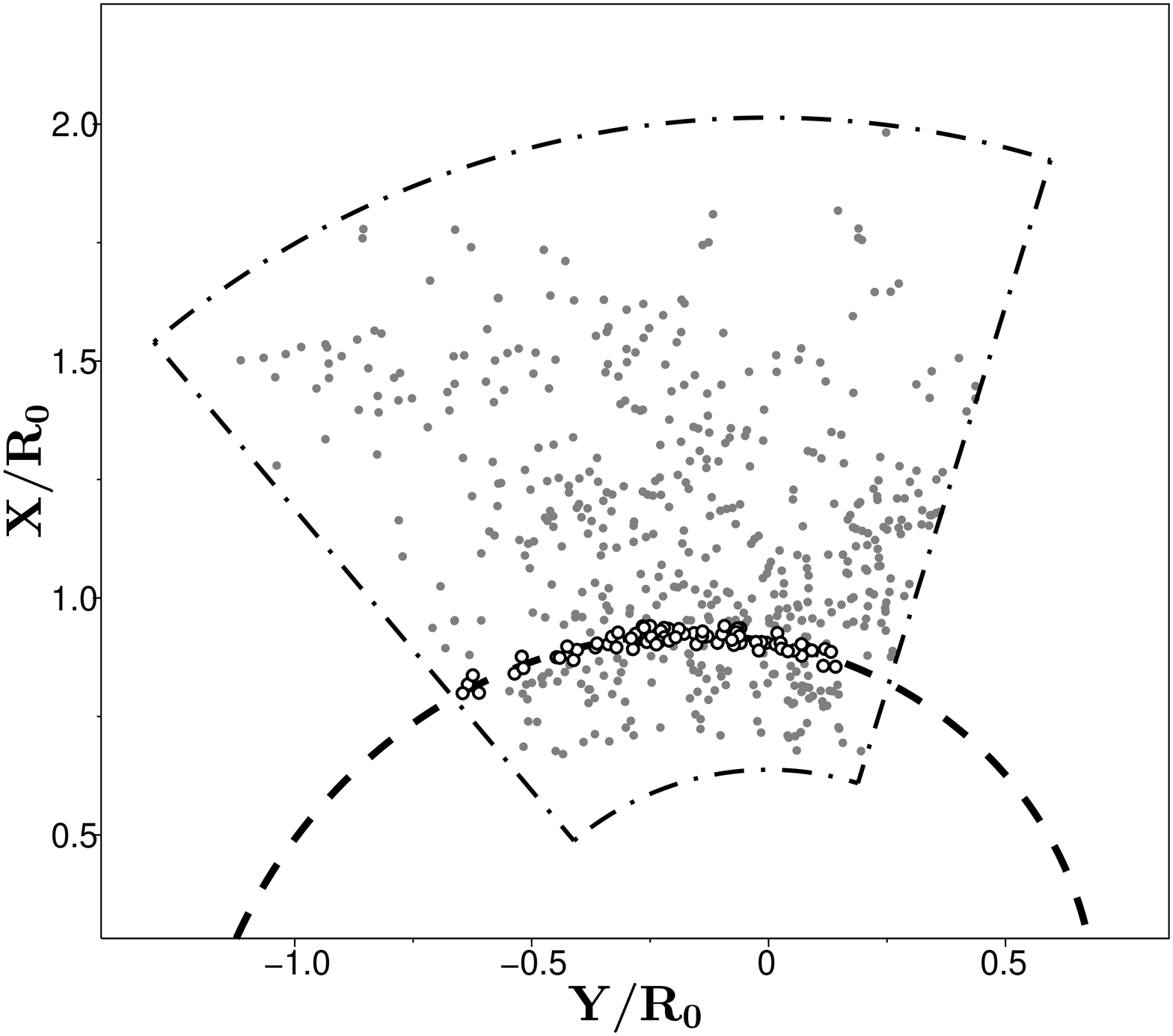}

\caption{\rm }
\label{xygcar}
\end{figure}

Figures~\ref{xygcar} and~\ref{thetalgcar}  show the minimal-scatter chain of 76 Cepheids (open 
circles) in the the XY and $\theta$ -- log~($R_G/R_0$) coordinates, respectively. As is evident 
from these figures, the spiral traced by Eq.~(\ref{carina_arm}) coincides with what is usually 
referred to as the Carina-Sagitarrius arm.

\subsection*{Search for other arms structures}
\noindent
To find other arm structures present in the distribution of the Cepheids of our sample, we 
assume that the pitch angles of these structures do not differ drastically from that of the major arm 
identified in the previous subsection. To identify such structures, we plot the histogram of the quantity
$x$=log~$(R_G/R_0)$ + tan~$i_{(car-sgr)} \theta$ (Fig.~\ref{histspir}). In this 
histogram three dips can be seen at $x$=-0.275, +0.075, and +0.325, which separate four 
peaks at $x~\sim$~-0.325, -0.100, +0.175, and +0.425. The most conspicuous among them
is the peak at $x~\sim$~-0.100, which corresponds to the Carina-Sgr arm already identified.
Our next target is the spiral that lies beyond the Carina-Sagittarius arm
when observed from the Sun, i.e., closer to the Galactic center. To determine its parameters,
we fit regression equation~(\ref{lspiral}) to the data for 23 stars with $x<$~-0.275:

\begin{equation}
\alpha_0 = -0.320 \pm 0.005~\hspace{0.3cm}~{\rm and}~{\rm tan}~i = -0.177 \pm 0.017~\hspace{0.3cm}~(i = -10.02^o \pm 0.95^o).
\label{inner_arm}
\end{equation}

with the cross-spiral scatter of $\sigma~{\rm log} (R_G/R_0) = 0.021$.
The parameters of the other two arms are:

\begin{equation}
\alpha_0 = +0.199 \pm 0.005~\hspace{0.3cm}~{\rm and}~{\rm tan}~i = -0.139 \pm 0.022~\hspace{0.3cm}~(i = -7.91^o \pm 1.24^o)
\label{perseus_arm}
\end{equation}

with the cross-spiral scatter of $\sigma~{\rm log} (R_G/R_0) = 0.067$ (the Perseus arm) and

\begin{equation}
\alpha_0 = +0.428 \pm 0.008~\hspace{0.3cm}~{\rm and}~{\rm tan}~i = -0.181 \pm 0.022~\hspace{0.3cm}~(i = -10.26^o \pm 1.22^o)
\label{outer_arm}
\end{equation}
with the cross-spiral scatter of $\sigma {\rm log} (R_G/R_0) = 0.055$ (the Outer arm).
We plot all four arms and the corresponding Cepheids in Figs.~\ref{allspirxy} and~\ref{allspir}.

We summarize the parameters of the
identified arms in Table~\ref{tab1}.
Note that while the two arms closest to the Galactic center  are
narrow and well defined, the two external arms appear rather loose and flocculent -- this
might be due to  the fact that the two outer arms are located beyond the corotation radius, whereas
the two inner arms are inside the corotation. The inferred pitch angles for all the arms are quite 
consistent except for that of the Perseus arm,
which is smaller by about 2$\sigma$ than the overall value. 
Forcing the same pitch angle for all four arms
with weights inversely proportional to 1/($\sigma$~x$^2$) gives the results summarized in Table~\ref{tab2}.

Finally, if we assume that the spiral pattern of the Milky Way has a four-armed structure and 
force our solution to
represent four global grand-design arms with each being identical to the previous one rotated by 90 degrees 
then the
global pitch angle and parameters $\alpha_0$ and $a_0$ are equal to the values listed in Table~\ref{tab3}. 
We show the
resulting grand-design spiral pattern in Fig.~\ref{grand}.
Note that the global pitch angle in this case ($i$=-9.46$^o$~$\pm$~0.11$^o$) is, on the whole, consistent
with the estimates obtained for individual arm fragments ($i$=-7.9$^o$--~-10.5$^o$, see Table~1). 
Fitting the arm fragments identified to a two- or three-armed global pattern yields global pitch
angles that are smaller in absolute value than the individual pitch angles of each of the four
fragments  ($i$=-4.59$^o$~$\pm$~0.19$^o$ and $i$=-6.68$^o$~$\pm$~0.15$^o$ for the two-
and three-armed pattern, respectively), and the cross-arm scatter of Cepheids ($\sigma$~x) 
relative to the corresponding spirals increases by a factor of  $\sim$~2 and $\sim$~2.3 
compared to the four-armed pattern, respectively. Moreover, the pitch angle determined
under the assumption of the three-armed spiral structure is at the lowest  limit of
the distribution of observed pitch angles for the nearest spiral galaxies [$|i|_{min} \sim$~6.5$^o$ -
see Kennicutt and Hodge (1982) and Kendall, Clarke, and Kennicutt (2015)], and the
''two-armed'' pitch angle falls outside the observed distribution. Note also that
three-armed spiral galaxies are very rare. Taken together, these circumstances can be viewed as 
indirect evidence for the four-armed structure of the spiral pattern in our Galaxy.

Based on the results listed in Table~\ref{tab3} we find the solar phase angle
in the spiral pattern to be 
$\chi_{odot}$~=~360*$\alpha_{Car-Sgr}$/($\alpha_{Per}$-$\alpha_{Car-Sgr}$)~=~-121$^o$~$\pm$~3$^o$.
Our arm pitch-angle estimates agree well with the recent results  of Bobylev \& Bajkova (2014a)
($i$ =-9.3$^o$--~-14.8$^o$) based on an analysis of the spatial distribution of Galactic masers.
The fact that our mean estimates are somewhat smaller  compared to those of Bobylev \& 
Bajkova (2014a) can be explained by larger ages of Cepheids (about 66~Myr on the average)
compared to those of masers: as time passes, stars born in the arm ''stretch'' along Galactocentric 
azimuth because the angular velocity of Galactic rotation decreases with Galactocentric radius.
The difference between our solar-phase estimate  ($\chi_{odot}$~=~-121$^o$~$\pm$~3$^o$)
from that of Bobylev \& Bajkova (2014a) ($\chi_{odot}$~=~-140$^o$~$\pm$~3$^o$) can be explained
by the shift of the Cepheid-delineated arms in the solar neighborhood relative to the spiral-wave
crests  (which can be assumed to be practically coincident with maser-delineated arms) 
because the rotation of the spiral pattern  lags behind the circular rotation
in the solar neighborhood: $\Omega(R_0)$~=~29.97~km/s/kpc is appreciably greater 
than the pattern speed $\Omega_P$~=~$\sim$~25~km/s/kpc (see the next Section).

\section*{PATTERN SPEED}
\noindent
We now try to estimate the angular speed $\Omega_{sp}$ of the Galactic spiral pattern. We do it by comparing
the positions of Cepheids in the  spiral arms with the positions of the corresponding arms traced by 
very young objects -- Galactic masers with accurate VLBI trigonometric parallaxes as determined
by Bobylev \& Bajkova~(2014a) (see Table~2 in their paper):
\begin{equation}
log (R_0/R_G)= \alpha_{maser} + {\rm tan}(i_{maser})\theta.
\label{maser}
\end{equation}
We begin with the best-determined Carina--Sagittarius arm. Its parameters determined from masers
are $\alpha_{maser}$ = -0.166 and tan $i_{maser}$ = -0.163.
Given the very young age of maser sources we assume that eq.~(\ref{maser}) defines the current location of 
the zero-age spiral arm 
(i.e., the locus where new stars are currently born). We can rewrite eq.~(\ref{maser}) in the following 
reversed form:
\begin{equation}
\theta = (log (R_0/R_G) - \alpha_{maser})/{\rm tan}(i_{maser}).
\label{maser2}
\end{equation}
We then account for the shift (azimuthal turn) of the spiral wave during time $t$ equal to the age of a 
particular Cepheid and find
that at the time of its birth the angle $\theta$ of the Cepheid is given  by the equation
\begin{equation}
\theta = [(log (R_0/R_G) - \alpha_{maser})/{\rm tan}(i_{maser})] - \Omega_P t.
\label{cepheid1}
\end{equation}
Finally, during time $t$ differential Galactic rotation should have increased angle $\theta$ 
by $\Omega(R_G) t$ and hence
the current coordinate $\theta$ of the Cepheid should be equal to:
\begin{equation}
\theta = [(log (R_0/R_G) - \alpha_{maser})/{\rm tan}(i_{maser})] - \Omega_P t + \Omega(R_G) t.
\label{cepheid2}
\end{equation}
It follows from this that
\begin{equation}
\Omega_P t = -\theta + [(log (R_0/R_G) - \alpha_{maser})/({\rm tan}(i_{maser}))]  + \Omega(R_G) t.
\label{cepheid3}
\end{equation}
We now compute the ages $t$ of 76 Carina--Sagittarius arm Cepheids based on their pulsation periods using
the period-age relations for fundamental-mode and first-overtone 
solar-metallicity ($Z$=0.02) Cepheids determined by Bono et al.~(2005) (see Table~4 in
the corresponding paper) and use the rotation-curve parameters determined by 
Bobylev \& Bajkova~(2014b) (eq.~(9) in the corresponding
paper) to compute the $\Omega(R_G)$ values. To determine $\Omega_P$, we solve eq.~(\ref{cepheid3})
via least squares method to obtain:
\begin{equation}
\Omega_P(Car-Sgr) = 26.0~\pm~0.4 \textup{ km/s/kpc}.
\label{omegacarsgr}
\end{equation}
We similarly determine the pattern-speed estimates based on the positions of Cepheids in the Perseus 
($\alpha_{maser}$ = +0.200 and tan $i_{maser}$ = -0.207) and Outer ($\alpha_{maser}$ = +0.563 and 
tan $i_{maser}$ = -0.214) arms
assuming that the rotation curve determined by the parameters listed in eq.~(9) in 
Bobylev \& Bajkova~(2014b) becomes 
flat beyond $R_G~>1.15 R_0$:
$$
\Omega_P(Per) = 24.5~\pm~0.4 \textup{ km/s/kpc}
$$

and

\begin{equation}
\Omega_P(Outer) = 25.0~\pm~0.5 \textup{ km/s/kpc}.
\label{omegaperseus}
\end{equation}
The three estimates agree excellently with each other, however, the similarly inferred result for the Inner arm:
\begin{equation}
\Omega_P(Inner) = 7~\pm~3 \textup{ km/s/kpc}.
\label{omegainner}
\end{equation}
appears highly discrepant, possibly due to (1) the fact that the second-order expansion fails to properly describe 
the behavior of $\Omega (R_G)$ at small Galactocentric distances and (2) the wrong delineation of the Inner arm 
in Bobylev \& Bajkova~(2014a), where it is based on only four masers. We therefore average the estimates 
obtained from Cepheids in the other three arms:
\begin{equation}
<\Omega_P>) = 25.2~\pm~0.5 \textup{ km/s/kpc}.
\label{omegamean}
\end{equation}
his result is consistent with recent independent determinations, which, according to the review by 
Gerhard~(2011)
mostly lie in the  20--30~km/s/kpc interval, and agrees well with the estimates obtained by 
Dias \& L\'epine (2005) -- $\Omega_{sp}$~=~26--26~km/s/kpc -- by  backward circular rotation (the method 
employed in this paper) for 599 clusters
and by integrating the full orbits backwards for 212 clusters with complete kinematical data.
and also with the recent estimate obtained by Junqueira et al. (2015)
based on an analysis of the correlation between the change of the energy and angular momentum 
of open clusters induced by the spiral density wave: $\Omega_P$~=~23.0~$\pm$~0.5~km/s/kpc.

\section*{CONCLUSIONS}
\noindent
Our analysis of the spatial distribution of 565 classical Cepheids based mostly on our extensive
photometric data to identify four spiral-arm fragments of the global spiral pattern.
Note that the outer arm, which is most distant from the Galactic center, is actually identified 
by the positions of newly discovered Cepheids. We found a well-defined signature of the
Carina--Sagittarius arm with 76 Cepheids out of 565 lying along a narrow spiral with a pitch angle of about
10.5$^{\rm o}$ and a cross-arm scatter of about 80~pc. We also found a rather sharp inner arm delineated 
by 23 Cepheids
with practically the same pitch angle ($\sim$~10.0$^{\rm o}$) and a cross-arm scatter ($\sim$~100~pc). The 
two spiral concentrations lying beyond
the solar circle (the Perseus and Outer arms) appear broader and more flocculent with the pitch angles 
of $\sim$~8.0$^{\rm o}$ 
and $\sim$~10.9$^{\rm o}$, respectively, and cross-arm scatter of $\sim$~600~pc. The 
four arms identified can be fitted to a four-armed  grand-design
pattern with the pitch angle of $i$=9.5~$\pm$~0.1$^{\rm o}$. We also estimate the pattern 
speed by comparing 
our inferred positions of Cepheid-delineated arms with the arm positions determined by 
Bobylev \& Bajkova (2014a) based 
on the data
for Galactic masers, which are very young objects. Three arms out of four yield very consistent estimates 
in the vicinity
of $<\Omega_P>)$ = 25.2~$\pm$~0.5 km/s/kpc, which agrees with most of other recent determinations. The 
estimate based on the
positions of the Inner arm is highly discrepant, possibly due to the inapplicability of the adopted second-
order expansion for the rotation curve and the fact that Bobylev \& Bajkova (2014a)
delineated the run of this arm
based only on four masers.

\section{Acknowledgments}

This work was supported in part by the Russian Foundation
for Basic Research (project nos.  13-02-00203 and 14-02-00472). 
Calculations of the light-curve parameters and
gamma-velocities of Cepheids were supported by Russian Scientific
Foundation grant no. 14-22-00041.

\pagebreak

\clearpage

\begin{table}[t]
\vspace{6mm}
\centering
{{\bf Table 1.} Parameters of spiral arms outlined by Cepheids}\label{tab1}

\vspace{5mm}\begin{tabular}{lrrrrrr}
\hline
  Arm &  $\alpha_0$ & tan~$i$ & $i$,  & $\sigma$~x & $a_0$ & $N$\\
      &             &         & degrees  &                            & kpc     &  \\
  \hline
Inner   & -0.320 & -0.177 & -10.02 & 0.021 & 5.16  & 23\\
        & $\pm$  0.005 & $\pm$  0.017 &  $\pm$  0.95 &       & $\pm$ 0.03 &  \\
\hline
Car-Sgr & -0.097 & -0.184 & -10.45 & 0.013 & 6.44 & 76\\
        & $\pm$  0.002 & $\pm$  0.007 &  $\pm$  0.41 &     &  $\pm$  0.02 & \\
\hline
Perseus & +0.199 & -0.139 &  -7.91 & 0.067 & 8.66 & 168\\
        &  $\pm$ 0.005 &  $\pm$ 0.022 &  $\pm$ 1.24 &  & $\pm$ 0.05     &   \\
\hline
Outer   & +0.428 & -0.181 & -10.26 & 0.055 & 10.89 & 99\\
        &  $\pm$ 0.008 &  $\pm$ 0.022 &  $\pm$ 1.22 &  &  $\pm$ 0.09     &   \\
\hline
Mean    &        & -0.180 & -10.20 &       &   \\
        &        &  0.006 &   0.34 &       &   \\
\hline
\end{tabular}
\end{table}

\begin{table}[t]

\vspace{6mm}
\centering
{{\bf Table 2.} Parameters of spiral arms outlined by Cepheids determined forcing the same pitch angle for all arms}\label{tab2}

\vspace{5mm} \begin{tabular}{lrr }
\hline
Arm &  $\alpha_0$ &  $a_0$ \\
      &             &  kpc   \\
  \hline
Inner   & -0.320  & 5.16 \\
        & $\pm$  0.005 & $\pm$ 0.03  \\
\hline
Car-Sgr & -0.096 & 6.45 \\
        & $\pm$  0.002  &  $\pm$  0.02 \\
\hline
Perseus & +0.196  & 8.64 \\
        &  $\pm$ 0.005  & $\pm$ 0.05     \\
\hline
Outer  & +0.443  & 11.06 \\
        &  $\pm$ 0.006 & $\pm$ 0.06    \\
\hline
tan($i$)   &  -0.177          &  \\
           &  $\pm$~0.007     &  \\
$i$, degrees &  -10.03     & \\
           &  $\pm$~0.37     &  \\
\hline
\end{tabular}
\end{table}

\begin{table}[t]

\vspace{6mm}
\centering
{{\bf Table 3.} Parameters of spiral arms outlined by Cepheids determined assuming that the
four arms identified represent the global four-armed grand-design spiral structure}\label{tab3}

\vspace{5mm} \begin{tabular}{lrr }
\hline
  Arm &  $\alpha_0$ &  $a_0$ \\
      &             &  kpc   \\
  \hline
Inner   & -0.349  & 5.01 \\
        & $\pm$  0.002 & $\pm$ 0.01  \\
\hline
Car-Sgr & -0.088 & 6.50 \\
        & $\pm$  0.002  &  $\pm$  0.01 \\
\hline
Perseus & +0.174  & 8.45 \\
        &  $\pm$ 0.002  & $\pm$ 0.02     \\
\hline
Outer   & +0.436  & 10.98 \\
        &  $\pm$ 0.006 & $\pm$ 0.02    \\
\hline
tan($i$)   &  -0.167          &  \\
           &  $\pm$~0.002     &  \\
$i$, degrees   &  -9.46     & \\
           &  $\pm$~0.11     &  \\
\hline
\end{tabular}
\end{table}

\clearpage

\centerline {\bf FIGURE CAPTIONS}
\vspace{1 cm}

Fig.1.~Distribution of classical Cepheids projected onto the Galactic plane. The big circle and
asterisk indicate the positions of the Galactic center and Sun, respectively.

Fig.2.~Distribution of classical Cepheids in in the
log ($R_g/R_0$), versus Galactocentric azimuth, $\theta$, coordinates.
The dashed rectangle shows the ''relative completeness'' region..

Fig.3.~Pitch angle $i$ of the minimal-scatter chain as a function of the number of stars $n$.

Fig.4.~Parameter $\alpha_0$ of the minimal-scatter chain as a function of the number of stars $n$..

Fig.5.~The largest minimum-scatter chain in the distribution of Cepheids in the XY coordinates.
The circles show the Cepheids that belong to the largest ($n$=76) chain and the
dashed line shows the corresponding logarithmic-spiral fit..

Fig.6.~The largest minimum-scatter chain in the distribution of Cepheids in the 
$\theta$ -- log~($R_G/R_0$)
coordinates. The circles show the Cepheids that belong to the largest ($n$=76) chain and the
dashed line shows the corresponding logarithmic-spiral fit..

\begin{figure}[ht]
\epsfxsize=13cm
\hspace{-2cm}\epsffile{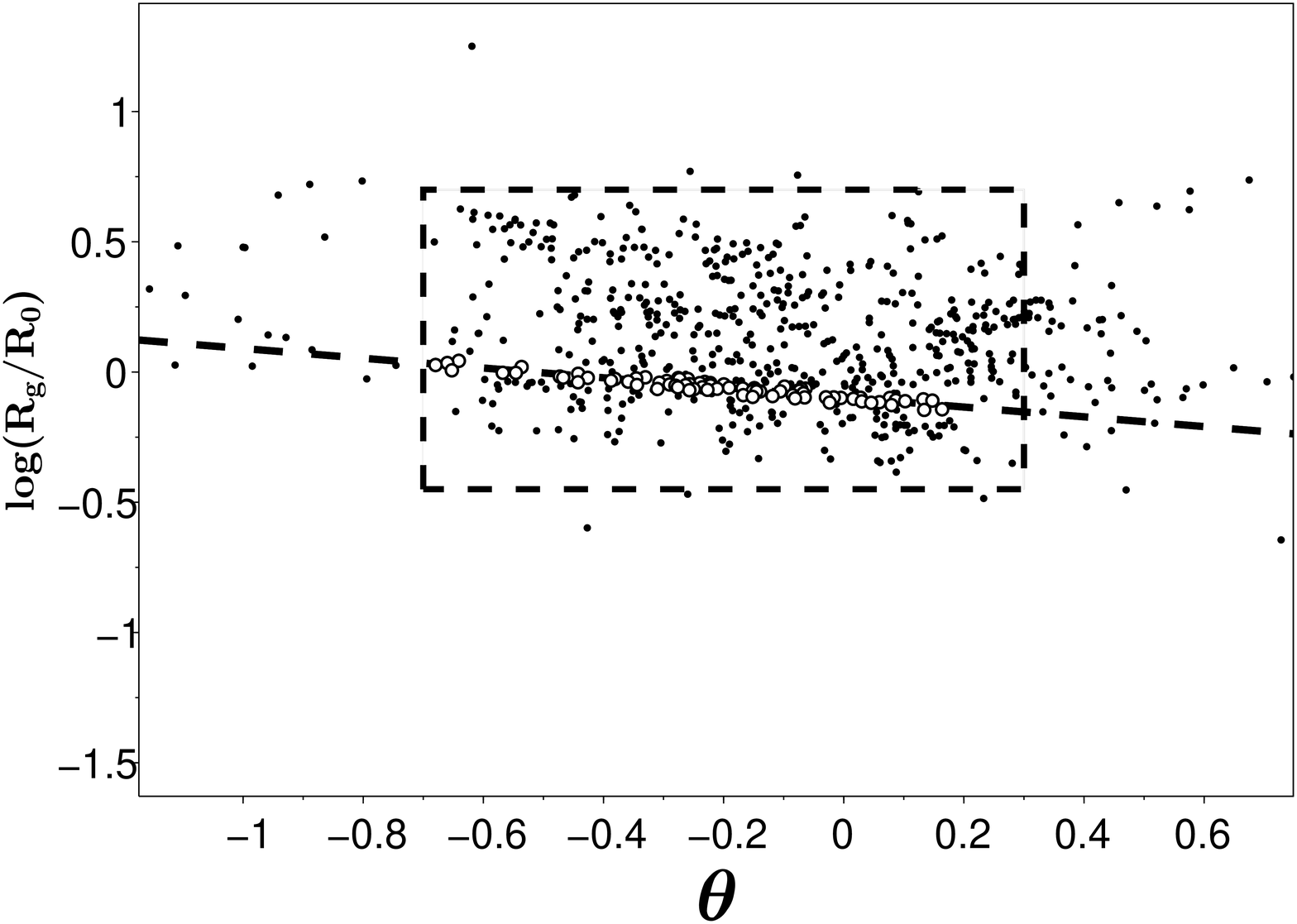}

\caption{\rm }
\label{thetalgcar}
\end{figure}

Fig.7.~Zoomed-in view of Fig~\ref{thetalgcar}.

Fig.8.~Histogram of $x = {\rm log} (R_G/R_0) + 0.184 \theta$ quantity.
The dips at $x$=-0.275, +0.075, and +0.325 separate the peaks at 
$x~\sim$~-0.325, -0.100, +0.175, and +0.425 that correspond to spiral arms.

Fig.9.~Positions of spiral arms projected onto the Galactic plane. The stars that outline
the Inner, Carina--Sagittarius, Perseus, and Outer arms are shown by the open circles, filled circles, triangles, and 
diamonds, respectively.

Fig.10.~Positions of spiral arms in the log ($R_g/R_0$) - $\theta$ coordinates. Designations
are the same as in Fig.~\ref{allspirxy}.

Fig.~11.~The grand-design spiral pattern based on the parameters listed in Table~\ref{tab3}.
The solid lines show the arm segments located within the area studied
(-0.7~$\leq$~$\theta$~$\leq$~+0.3 and -0.45~$\leq$~log($R_g/R_0)$~$\leq$~+0.7).

\clearpage

\begin{figure}[ht]
\epsfxsize=13cm
\hspace{-2cm}\epsffile{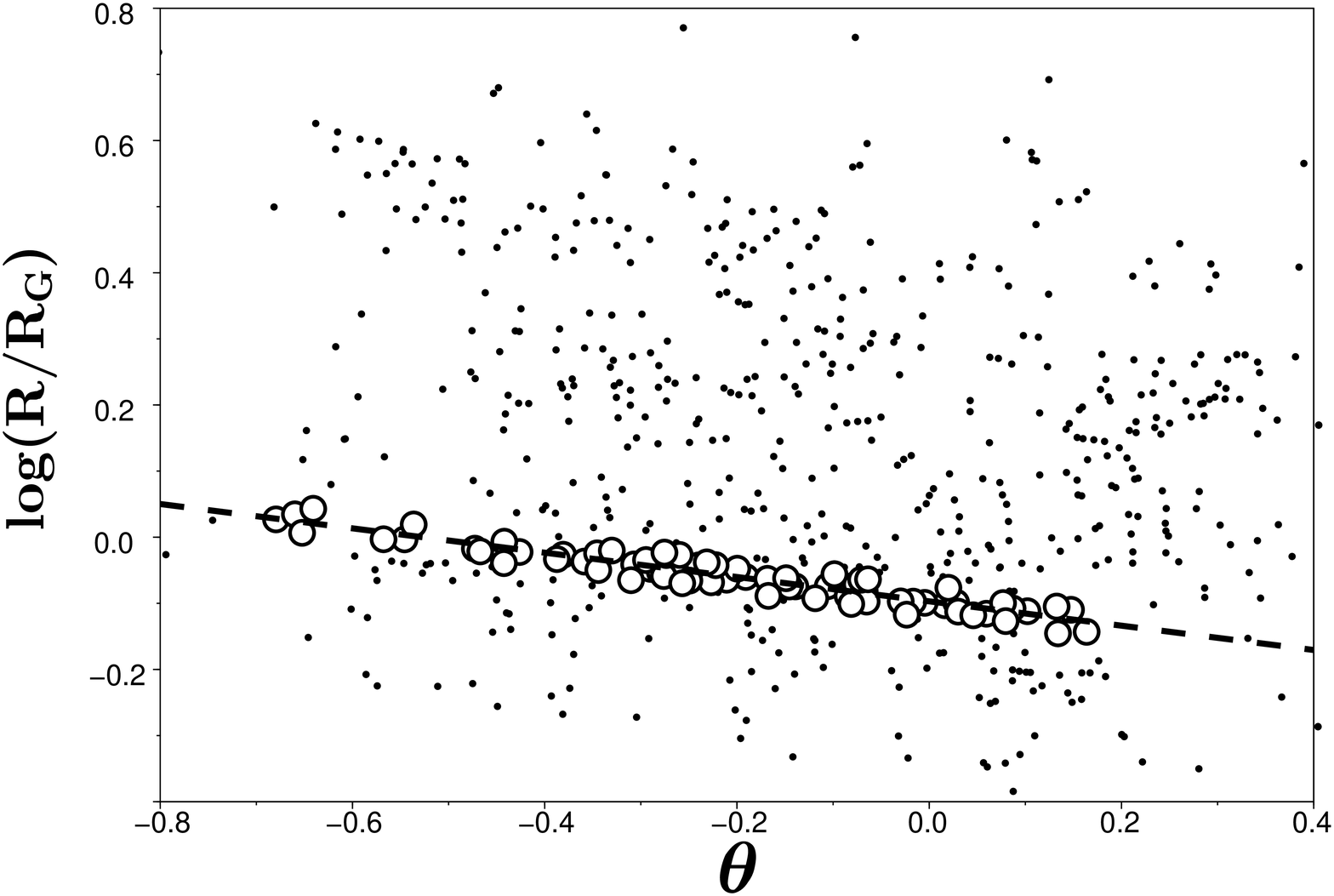}

\caption{\rm }
\label{thetalgcarnew}

\end{figure}

\begin{figure}[ht]
\epsfxsize=13cm
\hspace{-2cm}\epsffile{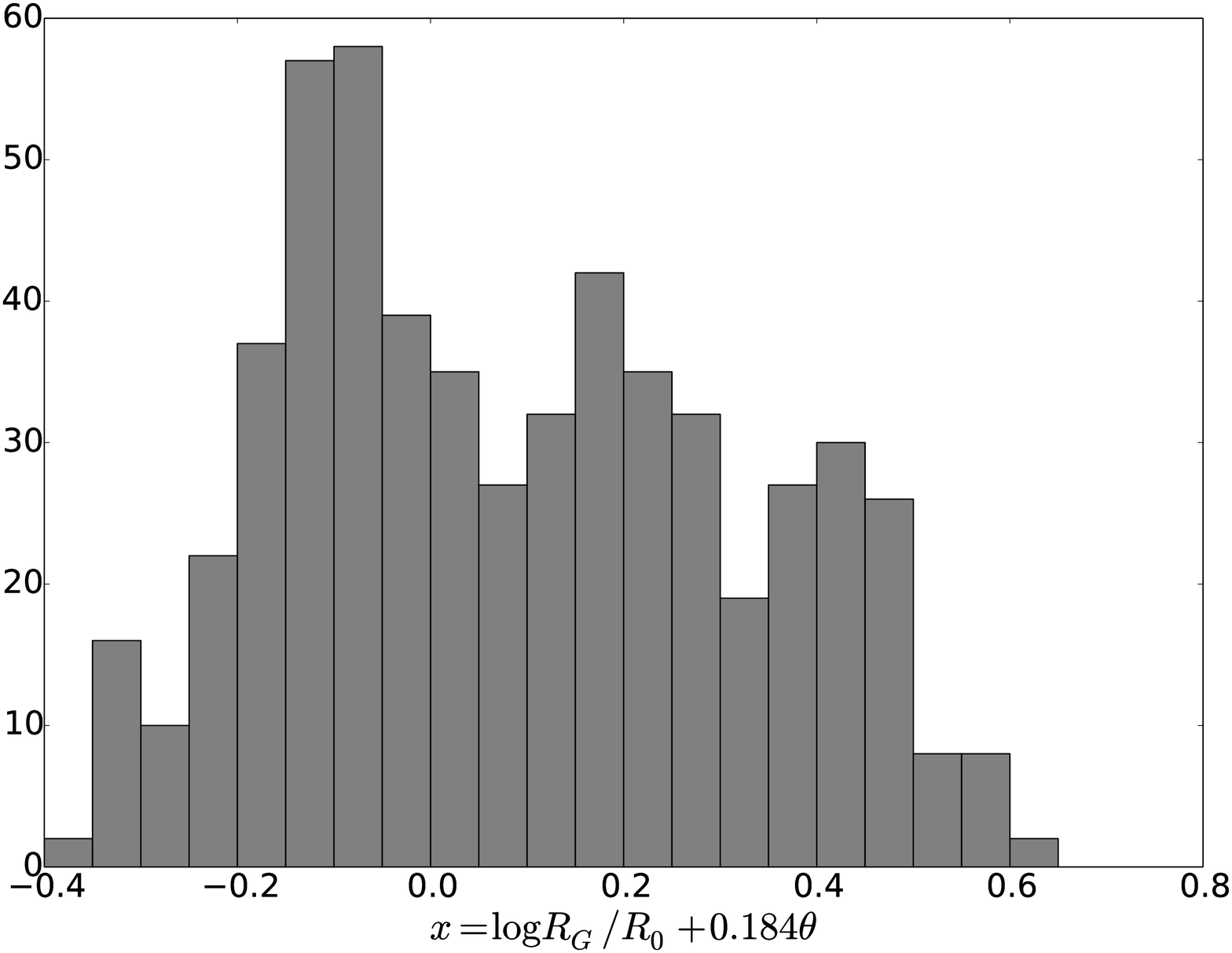}

\caption{\rm }
\label{histspir}
\end{figure}


\begin{figure}[ht]
\epsfxsize=13cm
\hspace{-2cm}\epsffile{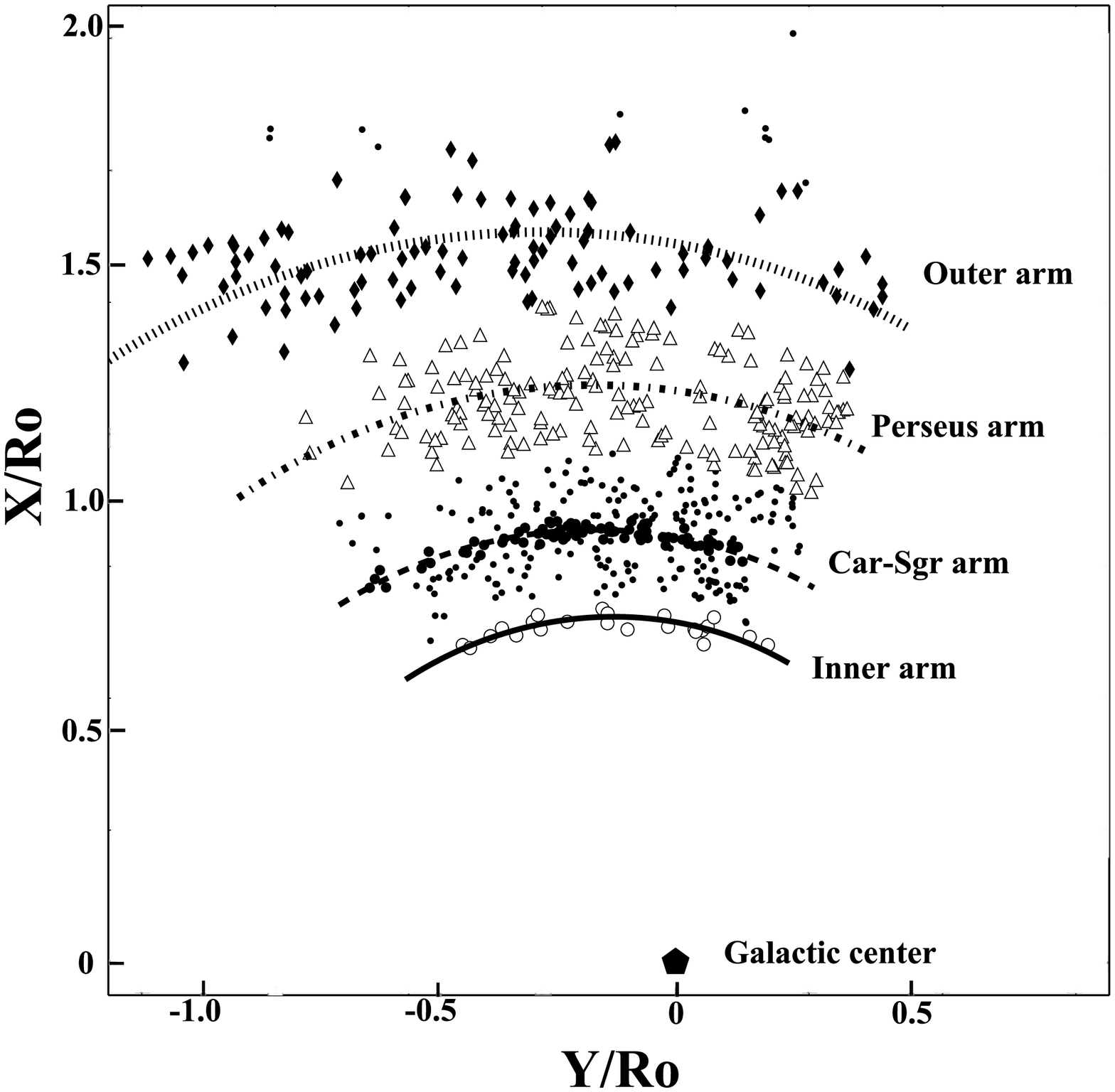}

\caption{\rm }
\label{allspirxy}
\end{figure}

\begin{figure}[ht]
\epsfxsize=13cm
\hspace{-2cm}\epsffile{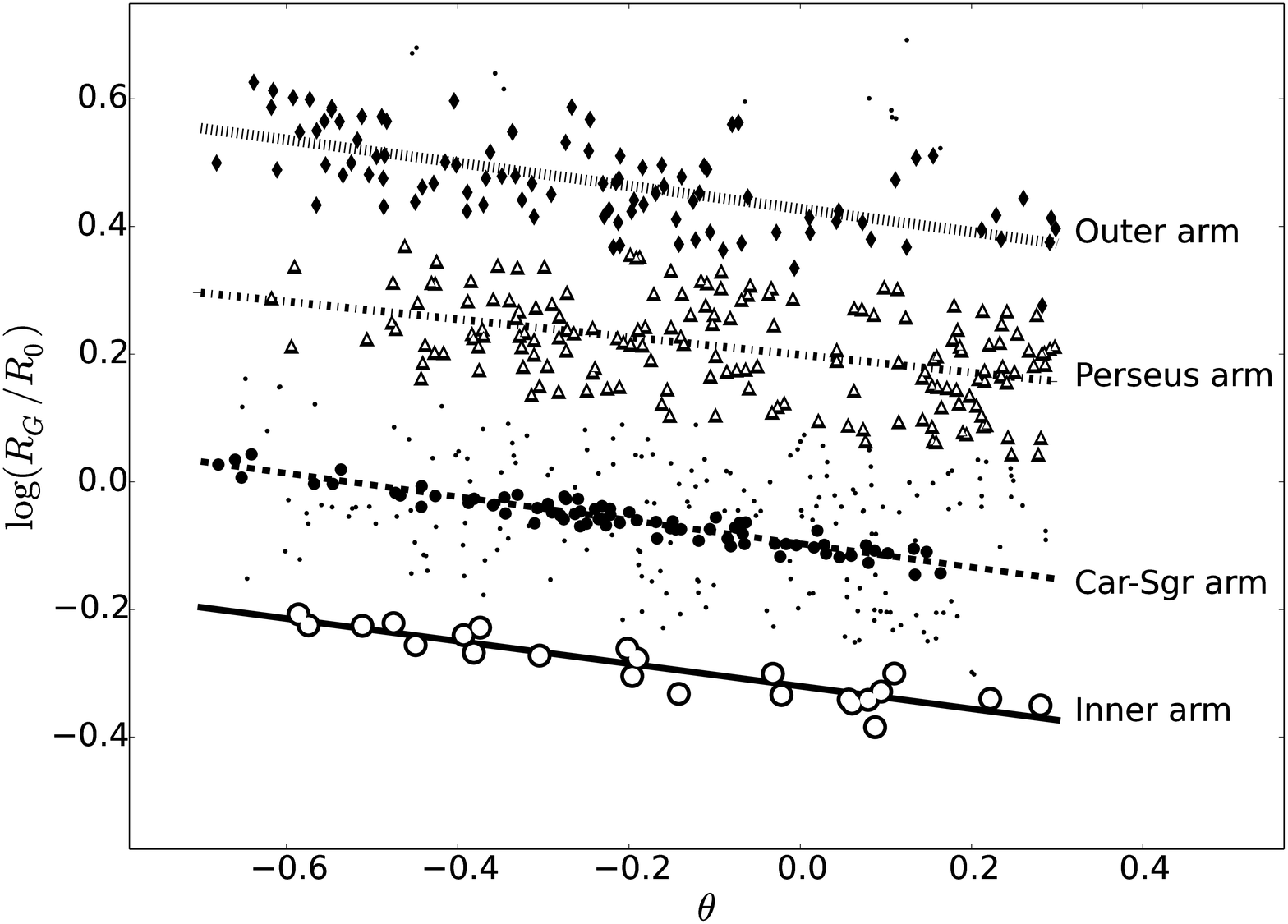}

\caption{\rm }
\label{allspir}
\end{figure}

\begin{figure}[ht]
\epsfxsize=13cm
\hspace{-2cm}\epsffile{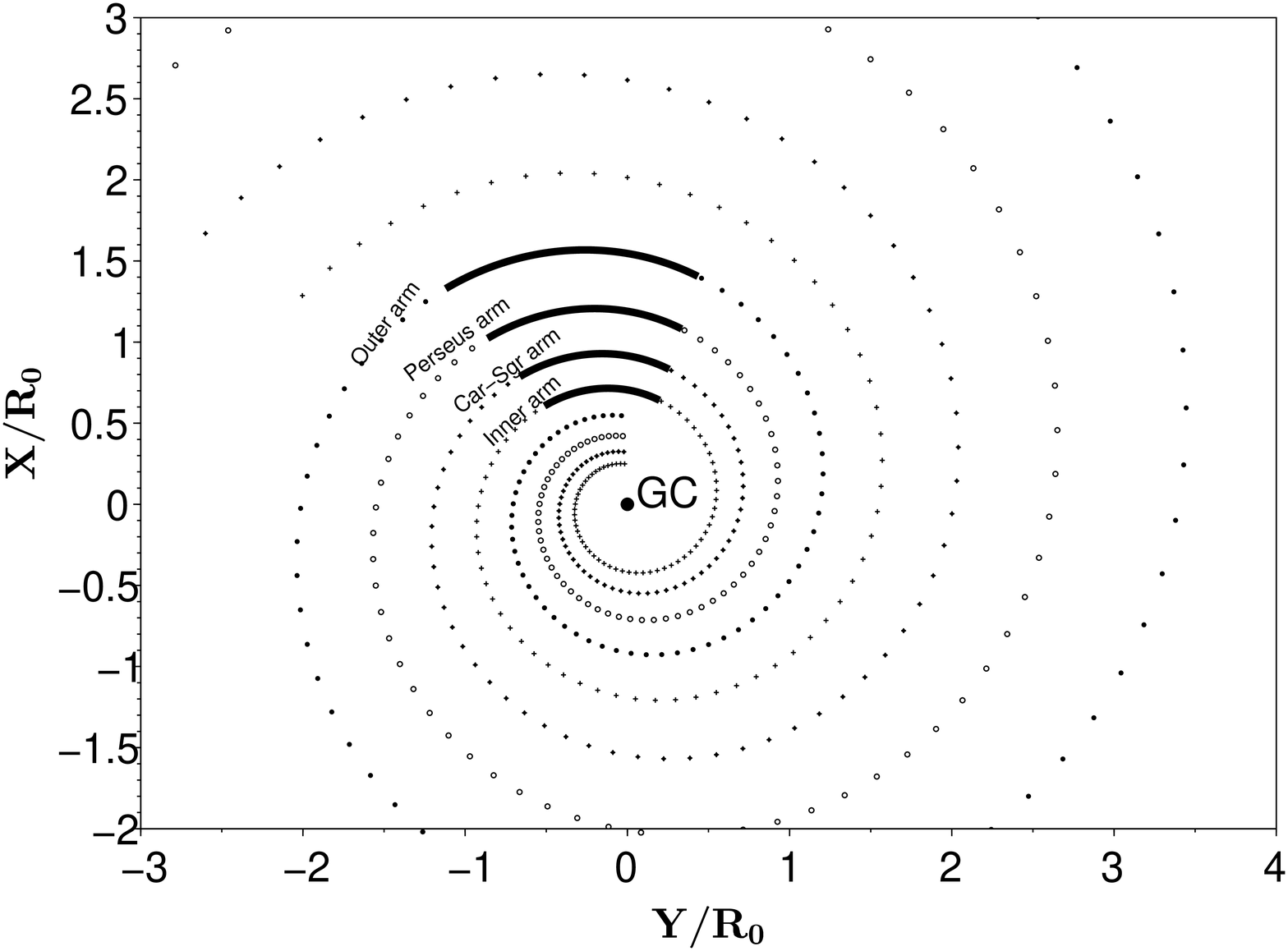}

\caption{\rm }
\label{grand}
\end{figure}


\begin{references}

\reference{\it
L.N.Berdnikov}, Odessa Astronomical Publications {\bf 18}, 23 (2006).

\reference{\it
L.N.Berdnikov, O.V.Vozyakova, A.K.Dambis}, Pis'ma Astron. Zh.
{\bf 22}, 936 (1996)


\reference{\it
L.N.Berdnikov}, VizieR On-line Data Catalog: CDS II/285 (2008).

\reference{\it
L.N.Berdnikov, A.K.Dambis, O.V.Vozyakova}, Astron. Astrophys. Suppl. {\bf 143}, 211 (2000).

\reference{\it
L.N.Berdnikov, A.Y.Kniazev, V.V.Kravtsov, E.N.Pastukhova, D.G.Turner}, Pis'ma Astron. Zh.
{\bf 35}, 44 (2009a)

\reference{\it
L.N.Berdnikov, A.Y.Kniazev, V.V.Kravtsov, E.N.Pastukhova, D.G.Turner}, Pis'ma Astron. Zh.
{\bf 35}, 348 (2009b)

\reference{\it
L.N.Berdnikov, A.Y.Kniazev, R. Sefako, V.V.Kravtsov, E.N.Pastukhova, S.V.Zhuiko},
\azh\ {\bf 88}, 886 (2011).

\reference{\it
L.N.Berdnikov, A.Y.Kniazev, R. Sefako, V.V.Kravtsov,  S.V.Zhuiko},
Pis'ma Astron. Zh. {\bf 40}, 147 (2014).


\reference{\it
L.N.Berdnikov, A.Y.Kniazev, R. Sefako, A.K.Dambis, V.V.Kravtsov,  S.V.Zhuiko},
Pis'ma Astron. Zh. {\bf 41}, 27 (2015).

\reference{\it
V.V.Bobylev, A.T.Bajkova} Pis'ma Astron. Zh. {\bf 39}, 899 (2013)

\reference{\it
V.V.Bobylev, A.T.Bajkova}, \mnras\ {\bf 437}, 1549 (2014a)


\reference{\it
V.V.Bobylev, A.T.Bajkova} Pis'ma Astron. Zh. {\bf 40}, 830 (2014b)

\reference{\it
G. Bono, M. Marconi, S. Cassisi, F. Caputo, W. Gieren, G. Pietrzynski}, \apj\ {\bf 621},
966 (2005)


\reference{\it
A. W. J. Cousins}, \memras\ {\bf 81}, 25 (1976)


\reference{\it
J. F. Dean, P.R. Warren, A. W. J. Cousins}, \mnras\ {\bf 183}, 569 (1978)

\reference{\it
W.S.Dias W. S., J. L\'epine}, \apj\ {\bf 629},
825 (2005)

\reference{\it
Yu. N. Efremov}, Astron. Zh. {\bf 55}, 127
(2011)

\reference{\it
P. Englmaier, O. Gerhard}, \mnras\ {\bf 304}, 512 (1999) 

\reference{\it
P. Englmaier,  M. Pohl, N. Bissantz},  Mem. Soc. Astron. Ital. Suppl. {\bf 18}, 199 (2011)


\reference{\it
P. Fouque, P.Arriagada, J.Storm, T.G.Barnes, N.Nardetto, A.Merand, P.Kervella, W.Gieren,
D.Bersier, G.F. Benedict, B.E.McArthur}, \aap\ {\bf 476}, 73 (2007)


\reference{\it
O.Gerhard}, Mem. Soc. Astron. Ital. Suppl. {\bf 18}, 185 (2011)

\reference{\it
C. Guerra, V. Pascucci}, IEICE Trans. Inf. \& Syst. {\bf E84-D}, 1739 (2001)

\reference{\it
L. G. Hou,  J. L. Han, W. B. Shi}, \aap\ {\bf 499}, 473 (2009)

\reference{\it
T. C. Junqueira, C. Chiappini, J. R. D. L\'epine, 
I. Minchev, B. X. Santiago}, \mnras\ {\bf 449}, 2336 (2015)

\reference{\it S. Kendall, 
C. Clarke, R. C. Kennicutt}, \mnras\ {\bf 446}, 4155 (2015)

\reference{\it R. Kennicutt, Jr.,
P. Hodge}, \apj\ {\bf 253}, 101 (1982)


\reference{\it
M. Kodric, A. Riffeser, U. Hopp, et al.}, \aj\ {\bf 145}, 106 (2013)

\reference{\it
Mi Kyoung Kim, Tomoya Hirota, Mareki Honma, et al.}, \pasj\ {\bf 60}, 991 (2008)

\reference{\it
E. S. Levine, L. Blitz, C. Heiles}, Science {\bf 312}, 1773 (2006)

\reference{\it
H. Nakanishi, Y. Sofue}, \pasj\ {\bf 55}, 191 (2003)

\reference{\it
M. Pohl, P. Englmaier, N. Bissantz} \apj\ {\bf 677}, 283 (2008)

\reference{\it
G. Pojmanski}, Acta Astron. {\bf 52}, 397 (2002)

\reference{\it
M.J.Reid, K.M. Menten, X.W.Zheng, A.Brunthaler, L. Moscadelli, Y. Xu, B. Zhang, M. Sato, M. Honma,
T. Hirota, K. Hachisuka, Y.K. Choi,  G.A. Moellenbrock, A. Bartkiewicz}, \apj\ {\bf 700}, 137 (2009)

\reference{\it
D. Russeil}, \aap\ {\bf 397}, 133 (2003)

\reference{\it
K. L. J. Rygl,  A. Brunthaler, M.J.Reid, K.M.Menten, H.J. van Langevelde, Y. Xu},
\aap\ {\bf 511}, 2 (2010)

\reference{\it
N.N.Samus, O.V.Durlevich, E.V.Kazarovets, N.N.Kireeva, E.N.Pastukhova, A.V.Zharova, et al.}, 
VizieR On-line Data Catalog: {\bf B/gcvs}, (2012)

\end{references}
\end{document}